# Spectral decomposition of 1D Fokker - Planck differential operator

*Igor A. Tanski*
*Moscow*
*povorot2@infoline.su*

V&T

*ABSTRACT*

We construct spectral decomposition of 1D Fokker - Planck differential operator. This reveal solution of Cauchy problem. We develop fundamental solution of Cauchy problem and compare it with one obtained by other means in our former work [5].

**1. Fokker - Planck 1D differential operator**

The Fokker - Planck equation in one dimension is

$$\frac{\partial n}{\partial t} + v \frac{\partial n}{\partial x} - \alpha \frac{\partial (vn)}{\partial v} - k \frac{\partial^2 n}{\partial v^2} = 0. \tag{1}$$

where

$n = n(t, x, v)$ - density;

$t$ - time variable;

$x$ - space coordinate;

$v$ - velocity;

$a$ - damping coefficient;

$k$ - diffusion coefficient.

(1) is evolutionary equation with differential operator

$$L(\phi) = -v \frac{\partial \phi}{\partial x} + \alpha \frac{\partial (v\phi)}{\partial v} + k \frac{\partial^2 \phi}{\partial v^2}. \tag{2}$$

where $\phi = \phi(x, v)$.

Differential operator (2) is called Fokker - Planck 1D differential operator. This operator is not self-adjoint. Its adjoint operator is:

$$L^*(\psi) = v \frac{\partial \psi}{\partial x} - \alpha v \frac{\partial \psi}{\partial v} + k \frac{\partial^2 \psi}{\partial v^2}. \tag{3}$$

We discuss simple cyclic boundary conditions

$$\phi(x + a, v) = \phi(x, v), \quad -\infty < v < \infty. \tag{4}$$

## 2. Eigenfunctions and eigenvalues of Fokker - Planck 1D differential operator

Eigenfunctions $\phi$ of differential operator $L$ must satisfy with differential equation

$$L(\phi) = \lambda\phi = -v\frac{\partial \phi}{\partial x} + \alpha \frac{\partial (v\phi)}{\partial v} + k\frac{\partial^2 \phi}{\partial v^2}. \tag{5}$$

and boundary conditions (4).

We search solution of equation (5) in the form

$$\phi = X(x)\ V(v). \tag{6}$$

The separation of variables provides two equations

$$X' = cX, \tag{7}$$

$$kV'' + \alpha(vV)' - cvV = \lambda V, \tag{8}$$

where $c = const$ is a separation constant.

The solution of (7) and (4) is

$$X_m = e^{2\pi i m \frac{x}{a}}; \tag{9}$$

where $m$ is integer and accordingly $c = 2\pi i \frac{m}{a}$.

We expect that solution $V$ of (8) satisfies with Fourier transform applicability conditions. The transformation is

$$V(v) = \int_{-\infty}^{+\infty} F(\omega)e^{i\omega t}d\omega, \tag{10}$$

and inversely is

$$F(\omega) = \frac{1}{2\pi}\int_{-\infty}^{+\infty} V(v)e^{-i\omega v}dv. \tag{11}$$

Transformation of equation (8) is

$$-k\omega^2 - \alpha\omega F' - icF' = \lambda F, \tag{12}$$

or

$$(\alpha\omega + ic)F' = -(\lambda + k\omega^2)F, \tag{13}$$

or finally

$$\frac{F'}{F} = -\frac{\lambda + k\omega^2}{\alpha\omega - 2\pi\frac{m}{a}}. \tag{14}$$

Let us denote

$$\omega_* = 2\pi\frac{m}{\alpha a}. \tag{15}$$

Then we have

$$\frac{F'}{F} = -\frac{k(\omega^2 - \omega_*^2) + \lambda + k\omega_*^2}{\alpha(\omega - \omega_*)} = -\frac{k}{\alpha}(\omega + \omega_*) - \frac{\lambda + k\omega_*^2}{\alpha(\omega - \omega_*)}. \tag{16}$$

Integration of (16) gives

$$\ln F = -\frac{k}{2\alpha}(\omega + \omega_*)^2 - \frac{\lambda + k\omega_*^2}{\alpha}\ln(\omega - \omega_*). \tag{17}$$

Let us denote

$$n = -\frac{\lambda + k\omega_*^2}{\alpha} \tag{18}$$

Take exponent of (17) and get

$$F = e^{-\frac{k}{2\alpha}(\omega+\omega_*)^2}(\omega - \omega_*)^n. \tag{19}$$

Form of expression (19) suggests, that $n$ should be nonnegative integer. Another possibilities we discuss later (see end of section 3). Eigenvalues of equation (8) are:

$$\boxed{\lambda = -\alpha n - k\left(\frac{2\pi m}{\alpha a}\right)^2.} \tag{20}$$

In order to perform inverse transform of (19), we use basic properties of Fourier transformation. First of all, we use wellknown fact

$$F(\omega) = e^{-a\omega^2} \to V(v) = \sqrt{\frac{\pi}{a}} e^{-\frac{v^2}{4a}}. \tag{21}$$

The second fact is shift theorem

$$F(\omega - a) \to e^{iav} V(v). \tag{22}$$

Thus we get transformation of the first multiplier in (19):

$$F = e^{-\frac{k}{2\alpha}(\omega+\omega_*)^2} \to V_{m0} = \left(2\pi \frac{\alpha}{k}\right)^{\frac{1}{2}} \exp(-i\omega_* v) \, e^{-\frac{\alpha}{2k}v^2} = \left(2\pi \frac{\alpha}{k}\right)^{\frac{1}{2}} \exp(-\frac{2\pi i m}{\alpha a} v) \, e^{-\frac{\alpha}{2k}v^2}. \tag{23}$$

These are eigenfunctions for maximum eigenvalue. Another way to get this solution we show in the next section.

The third fact is differentiation theorem

$$i\omega F \to \frac{dV}{dv}. \tag{24}$$

This implies, that

$$F = (\omega - \omega_*)^n e^{-\frac{k}{2\alpha}(\omega+\omega_*)^2} \to V = (-i)^n \left(\frac{d}{dv} - i\omega_*\right)^n \left[\left(2\pi \frac{\alpha}{k}\right)^{\frac{1}{2}} \exp(-\frac{2\pi i m}{\alpha a} v) \, e^{-\frac{\alpha}{2k}v^2}\right]. \tag{25}$$

We use usual convention, that operator acts on functions, staying on the right hand side of operator.

(25) gives eigenfunctions for the rest of eigenvalues. The aim of following calculations is to express them through Hermite polynomials.

For this purpose we firstly write differentiation operator $D$ in (25) in another way:

$$D = \frac{d}{dv} - i\omega_* = \exp(\frac{2\pi i m}{\alpha a} v) \left[\left(\frac{d}{dv}\right) \exp(-\frac{2\pi i m}{\alpha a} v)\right]. \tag{26}$$

According to before mentioned convention, this expression means, that we firstly multiply function by $\exp(-\frac{2\pi i m}{\alpha a} v)$, then we differentiate and at last we multiply result by $\exp(\frac{2\pi i m}{\alpha a} v)$. Two multipliers (the right and the left) have inverse values, therefore we can write arbitrary degree of $D$ in very simple form:

$$D^n = \exp(\frac{2\pi i m}{\alpha a} v)\left[\left(\frac{d}{dv}\right)^n \exp(-\frac{2\pi i m}{\alpha a} v)\right]. \tag{27}$$

Therefore we write (25) the following

$$V = (-i)^n \left(2\pi \frac{\alpha}{k}\right)^{\frac{1}{2}} \exp(\frac{2\pi i m}{\alpha a} v) \left(\frac{d}{dv}\right)^n \left[\exp(-\frac{4\pi i m}{\alpha a} v) e^{-\frac{\alpha}{2k}v^2}\right]. \tag{28}$$

Our next task is to transform exponent in (28)

$$\exp(-\frac{4\pi i m}{\alpha a} v) e^{-\frac{\alpha}{2k}v^2} = \exp\left[-\frac{\alpha}{2k}\left(v^2 + \frac{4\pi i m}{\alpha a}\frac{2k}{\alpha} v\right)\right] = \exp\left\{-\frac{\alpha}{2k}\left[\left(v + \frac{4\pi i m k}{\alpha^2 a}\right)^2 + \left(\frac{4\pi m k}{\alpha^2 a}\right)^2\right]\right\}. \tag{29}$$

This implies

$$V = (-i)^n \exp\left[-\frac{\alpha}{2k}\left(\frac{4\pi m k}{\alpha^2 a}\right)^2\right] \left(2\pi \frac{\alpha}{k}\right)^{\frac{1}{2}} \exp(\frac{2\pi i m}{\alpha a} v) \left(\frac{d}{dv}\right)^n \left\{\exp\left[-\frac{\alpha}{2k}\left(v + \frac{4\pi i m k}{\alpha^2 a}\right)^2\right]\right\}. \tag{30}$$

Hermite polynomials are defined as

$$H_n(z) = \frac{(-1)^n}{\rho} \frac{d^n}{dz^n} \rho; \tag{31}$$

where $\rho = e^{-z^2}$.

Add scaling factor and obtain from (31) expression for derivative:

$$\frac{d^n}{dz^n} e^{-az^2} = (-1)^n a^{n/2} e^{-az^2} H_n(\sqrt{a} z). \tag{32}$$

Substitute $a = \frac{\alpha}{2k}$ and get the following expression

$$V = (i)^n (\frac{\alpha}{2k})^{n/2} \exp\left[-\frac{\alpha}{2k}\left(\frac{4\pi m k}{\alpha^2 a}\right)^2\right] \left(2\pi \frac{\alpha}{k}\right)^{\frac{1}{2}} \exp(\frac{2\pi i m}{\alpha a} v) \exp\left[-\frac{\alpha}{2k}\left(v + \frac{4\pi i m k}{\alpha^2 a}\right)^2\right] H_n\left(\sqrt{\frac{\alpha}{2k}}\left(v + \frac{4\pi i m k}{\alpha^2 a}\right)\right). \tag{33}$$

Combine two exponents, drop constant multipliers and get final expression for eigenfunctions of differential operator (8):

$$V_{mn} = \exp(-\frac{2\pi i m}{\alpha a} v) \exp\left(-\frac{\alpha}{2k} v^2\right) H_n\left(\sqrt{\frac{\alpha}{2k}}\left(v + \frac{4\pi i m k}{\alpha^2 a}\right)\right). \tag{34}$$

Eigenfunctions of differential operator (5) are equal to

$$\phi_{mn} = \exp(2\pi i \frac{m}{a}(x - \frac{v}{\alpha})) \exp\left(-\frac{\alpha}{2k} v^2\right) H_n\left(\sqrt{\frac{\alpha}{2k}}\left(v + \frac{4\pi imk}{\alpha^2 a}\right)\right) \tag{35}$$

## 3. Eigenfunctions and eigenvalues of adjoint differential operator

Eigenfunctions $\psi$ of adjoint differential operator $L^*$ must satisfy with differential equation (see (3))

$$L^*(\psi) = \lambda \psi = v \frac{\partial \psi}{\partial x} - \alpha v \frac{\partial \psi}{\partial v} + k \frac{\partial^2 \psi}{\partial v^2}. \tag{36}$$

We perform substitution (7) once again. This time we use definition

$$Y_m = e^{-2\pi i m \frac{x}{a}}; \tag{37}$$

instead of (9). The difference is in the $m$ sign.

After variables separation (analogous to (6-7)) we have the following equation (analogous to (8)):

$$kW'' - \alpha v W' - 2\pi i \frac{m}{a} v W = \lambda W. \tag{38}$$

This operator is adjoint to (8). Note, that $m$ has another meaning, due to sign changing in (37). This fact is crucial for orthogonality analysis (see below).

Let us perform substitution

$$W = \exp\left(\frac{\alpha}{2k} v^2\right) V(v); \tag{39}$$

$$W' = \exp\left(\frac{\alpha}{2k} v^2\right) \left(V' + \frac{\alpha}{k} vV\right);$$

$$W'' = \exp\left(\frac{\alpha}{2k} v^2\right) \left(V'' + 2\frac{\alpha}{k} vV' + (\frac{\alpha}{k} v)^2 V + \frac{\alpha}{k} V\right)$$

in (38). Thus $V$ satisfies with (8). Therefore eigenfunctions of adjoint operator (38) are

$$W_{mn} = \exp\left(-\frac{2\pi im}{\alpha a} v\right) H_n\left(\sqrt{\frac{\alpha}{2k}}\left(v + \frac{4\pi imk}{\alpha^2 a}\right)\right) \tag{40}$$

Substitution (39) does not involve $\lambda$. Therefore eigenvalues of differential operator (5) and adjoint differential operator (37) are the same ( see (20)) in accordance with theory [1, 2].

Eigenfunction with maximum eigenvalue for given period $a$ is equal to

$$W_{m0} = \exp(-\frac{2\pi i m}{\alpha a} v) .\tag{41}$$

This solution we obtain in another way (not depending from Fourier transform), equating coefficients by degrees of $v$ in (38) to zero:

$$\alpha W'_{m0} + 2\pi i \frac{m}{a} W_{m0} = 0;\tag{42}$$

$$kW''_{m0} = \lambda W_{m0}.$$

These equations are compatible. The first equation gives solution (41) and the second gives expression (20) (with $n = 0$) for eigenvalue.

We get corresponding solution of (8) from (42), performing inverse substitution (39). Obviously, it coincides with (23).

Calculate commutator of two operators: operator $D$ from (26) and operator $L = k\left(\frac{d}{dv}\right)^2 + \alpha\left(\frac{d}{dv}\right)v - cv - \lambda$. We get easily

$$L(D(V)) - D(L(V)) = -\alpha D(V)\tag{43}$$

Therefore $L(V) = 0$ implies $(L + \alpha)(D(V)) = 0$.

This gives the rule for building of $L$ eigenfunction with eigenvalue $\lambda - \alpha$ strarting from known eigenfunction with eigenvalue $\lambda$.

Starting from basic solution (23) we get (28) by induction. This is another way to get expressions for eigenfunctions, independent from Fourier transform.

Another important differential operator is $D_- = \frac{d}{dv} + 2\pi i \frac{m}{\alpha a} + \frac{\alpha}{k} v$. This operator inversely builds eigenfunction of $L$ with eigenvalue $\lambda + \alpha$ from known eigenfunction with eigenvalue $\lambda$. We see, that for $n = 0$ must be $D_-(V_{m0}) = 0$ and we can find $V_{m0}$ from this equation.

We can represent $L$ as product of $D$ and $D_-$, analogous to Schroedinger factorization method. All theory is similar to the quantum mechanics linear oscillator theory.

Eigenfunctions of (36) are

$$\boxed{\psi_{mn} = \exp\left(-2\pi i \frac{m}{a}(x + \frac{v}{\alpha})\right) H_n\left(\sqrt{\frac{\alpha}{2k}}\left(v + \frac{4\pi i m k}{\alpha^2 a}\right)\right)}\tag{44}$$

We built all eigenfunctions of 1D Fokker - Planck differential operator. We can return to open question - which values can posess $n$ in expression for eigenvalues (20). In order to answer to this question we consider equation for eigenfunctions (38) of adjoint differential operator for $m = 0$. The point $v = 0$ is regular point of this equation, so we can search its solution in the form of Taylor series by degrees of $v$. We see, that result is $O(e^{\frac{\alpha}{k}v^2})$ unless $\lambda = -n\alpha$. If $\lambda = -n\alpha$ the series breaks and solution is Hermite polynomial.

## 4. Orthogonality of Eigenfunctions

It is wellknown, that eigenfunction of differential operator is orthogonal to eigenfunctions of adjoint differential operator, which belong to another eigenvalues. We check below this orthogonality for our special case.

Orthogonality of $V_{mp}$ and $W_{mq}$ follows from orthogonality of Hermite polynomials. The product $V\,W$ is equal to

$$V_{mp}\,W_{mq} = \exp(-\frac{4\pi im}{\alpha a}v)\exp\left(-\frac{\alpha}{2k}v^2\right)H_p\!\left(\sqrt{\frac{\alpha}{2k}}\left(v+\frac{4\pi imk}{\alpha^2 a}\right)\right)H_q\!\left(\sqrt{\frac{\alpha}{2k}}\left(v+\frac{4\pi imk}{\alpha^2 a}\right)\right) = \quad (45)$$

$$= \exp\!\left[-\frac{\alpha}{2k}\left(\frac{4\pi mk}{\alpha^2 a}\right)^2\right]\exp\!\left[-\frac{\alpha}{2k}\left(v+\frac{4\pi imk}{\alpha^2 a}\right)^2\right]H_p\!\left(\sqrt{\frac{\alpha}{2k}}\left(v+\frac{4\pi imk}{\alpha^2 a}\right)\right)H_q\!\left(\sqrt{\frac{\alpha}{2k}}\left(v+\frac{4\pi imk}{\alpha^2 a}\right)\right)$$

$$= \exp\!\left[-\frac{\alpha}{2k}\left(\frac{4\pi mk}{\alpha^2 a}\right)^2\right]\exp(-\xi^2)H_p(\xi)\,H_q(\xi)$$

where $\xi = \sqrt{\frac{\alpha}{2k}}\left(v+\frac{4\pi imk}{\alpha^2 a}\right)$.

We know (see [3], Theorem 54), that

$$I_{pq} = \int_{-\infty}^{\infty}\rho H_p(\xi)H_q(\xi)d\xi = \delta_{pq}2^p p!\sqrt{\pi}. \tag{46}$$

We apply parallel translation of integration path on complex plane and get:

$$\int_{-\infty}^{\infty}V_{mp}\,W_{mq}dv = \exp\!\left[-\frac{\alpha}{2k}\left(\frac{4\pi mk}{\alpha^2 a}\right)^2\right]\sqrt{\frac{2k}{\alpha}}\int_{-\infty}^{\infty}\exp(-\xi^2)H_p(\xi)\,H_q(\xi)d\xi = \tag{47}$$

$$= \exp\!\left[-\frac{\alpha}{2k}\left(\frac{4\pi mk}{\alpha^2 a}\right)^2\right]\sqrt{\frac{2\pi k}{\alpha}}\,\delta_{pq}2^p p!.$$

(47) proves, that eigenfunctions $V_{mp}$ and $W_{mq}$, which belong to the same $m$, are orthogonal. But $V_{pl}$ and $W_{qm}$ functions are not orthogonal, when $p \neq q$. Multipliers $X_p$ and $Y_q$ (see (7) and (37)) are responsible for orthogonality in this case.

$$\int_0^a X_p Y_q dx = a\delta_{pq}. \tag{48}$$

This proves orthogonality of $\phi_{mp}$ and $\psi_{nq}$

$$\int_0^a dx \int_{-\infty}^\infty \phi_{mp}\, \psi_{nq}\, dv = a\delta_{mn}\delta_{pq} \exp\left[-\frac{\alpha}{2k}\left(\frac{4\pi mk}{\alpha^2 a}\right)^2\right] \sqrt{\frac{2\pi k}{\alpha}}\, 2^p p!. \qquad (49)$$

We can write for arbitrary function $G(x, v)$ the following expansion

$$G(x, v) = \sum_{m=-\infty}^{m=\infty} \sum_{p=0}^{p=\infty} A_{mp}\phi_{mp}, \qquad (50)$$

where coefficients $A_{mn}$ are equal to

$$A_{mp} = \frac{1}{a} \exp\left[\frac{\alpha}{2k}\left(\frac{4\pi mk}{\alpha^2 a}\right)^2\right] \sqrt{\frac{\alpha}{2\pi k}}\, \frac{1}{2^p p!} \int_0^a dx \int_{-\infty}^\infty G(x, v)\, \psi_{mp}\, dv. \qquad (51)$$

## 5. Solution of Cauchy problem. Fundamental solution.

According to previous results, solution of Cauchy problem with initial condition $n_0 = G(x, v)$ is

$$n(t, x, v) = \sum_{m=-\infty}^{m=\infty} \sum_{p=0}^{p=\infty} \exp\left[-(\alpha p + k\left(\frac{2\pi m}{\alpha a}\right)^2)t\right] A_{mp}\phi_{mp}, \qquad (52)$$

where coefficients $A_{mp}$ are given by (51).

If $G(x, v) = \delta(x - \xi)\delta(v - u)$ then $A_{mp} = \frac{1}{a}\exp\left[\frac{\alpha}{2k}\left(\frac{4\pi mk}{\alpha^2 a}\right)^2\right]\sqrt{\frac{\alpha}{2\pi k}}\,\frac{1}{2^p p!}\psi_{mp}(\xi, u)$ and finally

$$\delta(x - \xi)\delta(v - u) = \frac{1}{a}\sqrt{\frac{\alpha}{2\pi k}} \sum_{m=-\infty}^{m=+\infty} \exp\left[\frac{\alpha}{2k}\left(\frac{4\pi mk}{\alpha^2 a}\right)^2\right] \sum_{p=0}^{p=\infty} \frac{\phi_{mp}(x, v)\psi_{mp}(\xi, u)}{2^p p!}; \qquad (53)$$

or

$$\delta(x - \xi)\delta(v - u) = \frac{1}{a}\sqrt{\frac{\alpha}{2\pi k}} \sum_{m=-\infty}^{m=+\infty} \exp\left(2\pi im\frac{(x-\xi)}{a}\right)\exp\left(2\pi im\frac{(v-u)}{\alpha a}\right)\times \qquad (54)$$

$$\times \sum_{p=0}^{p=\infty} \frac{1}{2^p p!} \exp\left[-\frac{\alpha}{2k}\left(v + \frac{4\pi imk}{\alpha^2 a}\right)^2\right] H_p\left(\sqrt{\frac{\alpha}{2k}}\left(v + \frac{4\pi imk}{\alpha^2 a}\right)\right) H_p\left(\sqrt{\frac{\alpha}{2k}}\left(u + \frac{4\pi imk}{\alpha^2 a}\right)\right).$$

In this way we obtain finally expansion of fundamental solution of Fokker - Planck equation:

$$G_a(t, x, \xi, v, u) = \frac{1}{a} \sqrt{\frac{\alpha}{2\pi k}} \sum_{m=-\infty}^{m=+\infty} \exp\left[-k\left(\frac{2\pi m}{\alpha a}\right)^2 t\right] \exp\left(2\pi im \frac{(x-\xi)}{a}\right) \exp\left(2\pi im \frac{(v-u)}{\alpha a}\right) \times \quad (55)$$

$$\times \sum_{p=0}^{p=\infty} \frac{\exp(-\alpha pt)}{2^p p!} \exp\left[-\frac{\alpha}{2k}\left(v + \frac{4\pi imk}{\alpha^2 a}\right)^2\right] H_p\left(\sqrt{\frac{\alpha}{2k}}\left(v + \frac{4\pi imk}{\alpha^2 a}\right)\right) H_p\left(\sqrt{\frac{\alpha}{2k}}\left(u + \frac{4\pi imk}{\alpha^2 a}\right)\right)$$

where $\xi$ - initial point of space;

$u$ - initial velocity.

Index $a$ below points to interval's length.

We can simplify (55) by explicit summation of the second sum. As an instrument for this task we use following identity found in [3] (Theorem 53) with reference to the original work [4].

$$\sum_{n=0}^{n=\infty} \frac{e^{-\frac{1}{2}(x^2+y^2)}}{2^n n!} s^n H_n(x) H_n(y) = \frac{1}{\sqrt{1-s^2}} \exp\left[\frac{x^2 - y^2}{2} - \frac{(x - ys)^2}{1 - s^2}\right]. \quad (56)$$

This identity is true when $|s| < 1$.

We perform substitution in (55)

$$\exp(-\alpha t) = s; \quad (57)$$

$$\sqrt{\frac{\alpha}{2k}}\left(v + \frac{4\pi imk}{\alpha^2 a}\right) = \mu;$$

$$\sqrt{\frac{\alpha}{2k}}\left(u + \frac{4\pi imk}{\alpha^2 a}\right) = \nu.$$

After that the second sum in (55) is equal to

$$\sum_{p=0}^{p=\infty} \frac{s^p}{2^p p!} \exp[-\mu^2] H_p(\mu) H_p(\nu) = \quad (58)$$

$$= \exp\left[\frac{-\mu^2 + \nu^2}{2}\right] \frac{1}{\sqrt{1-s^2}} \exp\left[\frac{\mu^2 - \nu^2}{2} - \frac{(\mu - \nu s)^2}{1-s^2}\right] = \frac{1}{\sqrt{1-s^2}} \exp\left[-\frac{(\mu - \nu s)^2}{1-s^2}\right].$$

or

$$G_a(t, x, \xi, v, u) = \frac{1}{a} \sqrt{\frac{\alpha}{2\pi k}} \sum_{m=-\infty}^{m=+\infty} \exp\left[-k\left(\frac{2\pi m}{\alpha a}\right)^2 t\right] \frac{\exp\left(2\pi im \frac{(x-\xi)}{a}\right)}{\sqrt{1 - e^{-2\alpha t}}} \times \quad (59)$$

$$\times \exp\left(2\pi im \frac{(v-u)}{\alpha a}\right) \exp\left[-\frac{(\mu - \exp(-\alpha t)\nu)^2}{1 - \exp(-2\alpha t)}\right].$$

We return from $\mu, \nu$ to $v, u$ variables

$$G_a(t, x, \xi, v, u) = \frac{1}{a}\sqrt{\frac{\alpha}{2\pi k}}\sum_{m=-\infty}^{m=+\infty}\exp\left[-k\left(\frac{2\pi m}{\alpha a}\right)^2 t\right]\frac{\exp\left(2\pi im\frac{(x-\xi)}{a}\right)}{\sqrt{1-e^{-2\alpha t}}}\exp\left(2\pi im\frac{(v-u)}{\alpha a}\right)\times \quad (60)$$

$$\times \exp\left[-\frac{\alpha}{2k(1-e^{-2\alpha t})}\left[(v-e^{-\alpha t}u) + (1-e^{-\alpha t})\frac{4\pi imk}{\alpha^2 a}\right]^2\right].$$

Expand square in the last exponent

$$G_a(t, x, \xi, v, u) = \frac{1}{a}\sqrt{\frac{\alpha}{2\pi k}}\sum_{m=-\infty}^{m=+\infty}\exp\left[-k\left(\frac{2\pi m}{\alpha a}\right)^2 t\right]\frac{\exp\left(2\pi im\frac{(x-\xi)}{a}\right)}{\sqrt{1-e^{-2\alpha t}}}\exp\left(2\pi im\frac{(v-u)}{\alpha a}\right)\times \quad (61)$$

$$\times \exp\left[-\frac{\alpha}{2k(1-e^{-2\alpha t})}\left[(v-e^{-\alpha t}u)^2 + 2(v-e^{-\alpha t}u)(1-e^{-\alpha t})\frac{4\pi imk}{\alpha^2 a} - (1-e^{-\alpha t})^2\left(\frac{4\pi mk}{\alpha^2 a}\right)^2\right]\right].$$

and then arrange arguments of exponents. This gives the final form of fundamental solution of initial value problem with simple cyclic boundary conditions (4):

$$G_a(t, x, \xi, v, u) = \frac{1}{a}\left(\frac{\alpha}{2\pi k(1-e^{-2\alpha t})}\right)^{\frac{1}{2}}\exp\left[-\frac{\alpha}{2k}\frac{(v-e^{-\alpha t}u)^2}{(1-e^{-2\alpha t})}\right]\times \quad (62)$$

$$\times \sum_{m=-\infty}^{m=+\infty}\exp\left[-k\left(\frac{2\pi m}{\alpha a}\right)^2\left(t - \frac{2}{\alpha}\frac{1-e^{-\alpha t}}{1+e^{-\alpha t}}\right)\right]\exp\left[\frac{2\pi im}{a}\left((x-\xi) + \frac{1}{\alpha}(v-u) - \frac{2}{\alpha}\frac{(v-e^{-\alpha t}u)}{(1+e^{-\alpha t})}\right)\right].$$

## 6. Fundamental solution for the whole space

Direct interval length $a$ to infinity and get from (62) fundamental solution for the whole space. The Fourier series sum is transformed to Fourier integral. Formally, we perform substitutions $\omega = \frac{2\pi m}{a}$ and $d\omega \approx \frac{2\pi}{a}$.

$$G_\infty(t, x, \xi, v, u) = \frac{1}{2\pi}\left(\frac{\alpha}{2\pi k(1-e^{-2\alpha t})}\right)^{\frac{1}{2}}\exp\left[-\frac{\alpha}{2k}\frac{(v-e^{-\alpha t}u)^2}{(1-e^{-2\alpha t})}\right]\times \quad (63)$$

$$\times \int_{-\infty}^{+\infty}\exp\left[-k\left(\frac{\omega}{\alpha}\right)^2\left(t - \frac{2}{\alpha}\frac{1-e^{-\alpha t}}{1+e^{-\alpha t}}\right)\right]\exp\left[i\omega\left((x-\xi) + \frac{1}{\alpha}(v-u) - \frac{2}{\alpha}\frac{(v-e^{-\alpha t}u)}{(1+e^{-\alpha t})}\right)\right]d\omega.$$

We perform calculations (63) using Gaussian integral

$$\frac{1}{2\pi} \int_{-\infty}^{+\infty} e^{-D\omega^2} e^{i\omega p} d\omega = \frac{1}{2\sqrt{\pi D}} e^{-\frac{p^2}{4D}}. \tag{64}$$

For our task $D$ and $p$ are equal to

$$p = (x - \xi) + \frac{1}{\alpha}(v - u) - \frac{2}{\alpha} \frac{(v - e^{-\alpha t} u)}{(1 + e^{-\alpha t})} ; \tag{65}$$

$$D = \frac{k}{\alpha^2} \left( t - \frac{2}{\alpha} \frac{1 - e^{-\alpha t}}{1 + e^{-\alpha t}} \right). \tag{66}$$

Use another variable $D_*$:

$$D_* = \frac{\alpha t \, (1 - e^{-2\alpha t}) - 2(1 - e^{-\alpha t})^2}{2\alpha^4}. \tag{67}$$

This variable was introduced in our former work [5] as determinant of quadratic form matrix. We have:

$$D = D_* \frac{2k\alpha}{(1 - e^{-2\alpha t})}. \tag{68}$$

Perform transformation and get another form of fundamental solution of Cauchy problem for the whole space:

$$G_\infty(t, x, \xi, v, u) = \left( \frac{\alpha}{2\pi k(1 - e^{-2\alpha t})} \right)^{\frac{1}{2}} \exp\left[ -\frac{\alpha}{2k} \frac{(v - e^{-\alpha t} u)^2}{(1 - e^{-2\alpha t})} \right] \times \tag{69}$$

$$\times \frac{1}{2} \left( \frac{(1 - e^{-2\alpha t})}{2\pi k\alpha D_*} \right)^{\frac{1}{2}} \exp\left\{ -\frac{1}{4kD_*} \frac{(1 - e^{-2\alpha t})}{2\alpha} \left( (x - \xi) + \frac{1}{\alpha}(v - u) - \frac{2}{\alpha} \frac{(v - e^{-\alpha t} u)}{(1 + e^{-\alpha t})} \right)^2 \right\}.$$

Join two exponents

$$G_\infty(t, x, \xi, v, u) = \left( \frac{\alpha}{2\pi k(1 - e^{-2\alpha t})} \right)^{\frac{1}{2}} \times \tag{70}$$

$$\times \frac{1}{2} \left( \frac{(1 - e^{-2\alpha t})}{2\pi k\alpha D_*} \right)^{\frac{1}{2}} \exp\left\{ -\frac{1}{4kD_*} \left[ \frac{(1 - e^{-2\alpha t})}{2\alpha} \left( (x - \xi) + \frac{1}{\alpha}(v - u) - \frac{2}{\alpha} \frac{(v - e^{-\alpha t} u)}{(1 + e^{-\alpha t})} \right)^2 + \frac{\alpha t \, (1 - e^{-2\alpha t}) - 2(1 - e^{-\alpha t})^2}{\alpha^3} \frac{(v - e^{-\alpha t} u)^2}{(1 - e^{-2\alpha t})} \right] \right\}.$$

or

$$G_\infty(t, x, \xi, v, u) = \frac{1}{4\pi k\sqrt{D_*}} \times \tag{71}$$

$$\times \exp\left\{-\frac{1}{4kD_*}\left[\frac{(1-e^{-2\alpha t})}{2\alpha}\left(\left((x-\xi) + \frac{1}{\alpha}(v-u)\right)^2 - \frac{4}{\alpha}\left((x-\xi) + \frac{1}{\alpha}(v-u)\right)\frac{(v-e^{-\alpha t}u)}{(1+e^{-\alpha t})}\right) + \frac{t}{\alpha^2}(v-e^{-\alpha t}u)^2\right]\right\}.$$

In order to perform final step, we introduce another set of variables:

$$\hat{x} = x - (\xi + \frac{u}{\alpha}(1 - e^{-\alpha t})); \tag{72}$$

$$\hat{v} = v - ue^{-\alpha t}.$$

These variables were introduced in [5] as integrals of characteristic equations. Subexpression of (71) is equal to

$$(x - \xi) + \frac{1}{\alpha}(v - u) = \hat{x} + \frac{\hat{v}}{\alpha}, \tag{73}$$

and whole (71) is now

$$G_\infty(t, x, \xi, v, u) = \frac{1}{4\pi k\sqrt{D_*}} \exp\left\{-\frac{1}{4kD_*}\left[\frac{(1-e^{-2\alpha t})}{2\alpha}\left(\left(\hat{x} + \frac{\hat{v}}{\alpha}\right)^2 - \frac{4}{\alpha}\left(\hat{x} + \frac{\hat{v}}{\alpha}\right)\frac{\hat{v}}{(1+e^{-\alpha t})}\right) + \frac{t}{\alpha^2}\hat{v}^2\right]\right\} \tag{74}$$

We see, that exponent's argument is quadratic form of new variables $\hat{x}, \hat{v}$.

Very simple calculation brings this expression to final form:

$$G_\infty(t, x, \xi, v, u) = \frac{1}{4\pi k\sqrt{D_*}} \exp\left\{-\frac{1}{4kD_*}\left[\frac{(1-e^{-2\alpha t})}{2\alpha}\hat{x}^2 - \right.\right. \tag{75}$$

$$\left.\left. - \left(\frac{2}{\alpha^2}(1-e^{-\alpha t}) - \frac{1}{\alpha^2}(1-e^{-2\alpha t})\right)\hat{x}\hat{v} + \left(\frac{t}{\alpha^2} - \frac{2}{\alpha^3}(1-e^{-\alpha t}) + \frac{1}{2\alpha^3}(1-e^{-2\alpha t})\right)\hat{v}^2\right]\right\}.$$

As expected, this result coincide with obtained in [5].

We checked our considerations, comparing result with earlier one obtained by other means.

**DISCUSSION**

The main result of these considerations is the closed form expression for eigenfunctions and eigenvalues of 1D Fokker - Planck differential operator. The spectral properties of 1D Fokker - Planck differential operator are very similar to properties of Hermitian quantum mechanics operators, especially - harmonic oscillator. All eigenvalues are real. For cyclic boundary conditions the spectrum is discrete, for the whole space the spectrum is continuous. Eigenvalues are nonpositive. The maximum (zero) eigenvalue corresponds to space uniform Maxwellian velocity distribution with zero mean velocity. Fundamental solution of Cauchy problem is obtained and compared with obtained in former work [5]. These results enable solution of more complex problems.

____________________________